\documentclass[superscriptaddress,preprintnumbers,amsmath,amssymb]{revtex4}
\usepackage{graphicx}% Include Figure files
\usepackage{dcolumn}% Align table columns on decimal point
\usepackage{bm}% bold math
\usepackage{latexsym}
\begin{document}
\title{\bf Molecular motors: design, mechanism and control} 
\author{Debashish Chowdhury}
\affiliation{Physics Department, Indian Institute of Technology, Kanpur 28016,\\and\\Max-Planck Institute for Physics of Complex Systems, Dresden, Germany}
\email{debch@iitk.ac.in}
\begin{abstract} 
Biological functions in each animal cell depend on coordinated operations 
of a wide variety of molecular motors. Some of the these motors transport 
cargo to their respective destinations whereas some others are mobile 
workshops which synthesize macromolecules while moving on their tracks. 
Some other motors are designed to function as packers and movers. All 
these motors require input energy for performing their mechanical works 
and operate under conditions far from thermodynamic equilibrium. The 
typical size of these motors and the forces they generate are of the 
order of nano-meters and pico-Newtons, respectively. They are subjected 
to random bombardments by the molecules of the surrounding aqueous medium 
and, therefore, follow noisy trajectories. Because of their small inertia, 
their movements in the viscous intracellular space exhibits features that 
are characteristics of hydrodynamics at low Reynold's number. In this 
article we discuss how theoretical modeling and computer simulations of 
these machines by physicists are providing insight into their mechanisms 
which engineers can exploit to design and control artificial nano-motors. 
\end{abstract}
\maketitle

%%%%%%%%%%%%%%%%%%%%%%%%%%%%%%%%%%%%%%%%%%%%%%%%%%%%%%%%%%%%%
\section{Introduction}
%%%%%%%%%%%%%%%%%%%%%%%%%%%%%%%%%%%%%%%%%%%%%%%%%%%%%%%%%%%%%

A cell, the structural and functional unit of life, in an animal body 
is not a passive bag of viscous multi-component fluid. The interior 
of a cell has lot of similarities with any modern city; just as there 
are streets, highways and railroad tracks on which traffic of vehicles 
move passengers and cargo in various destinations, intracellular 
molecular cargoes are also transported by wide varieties of molecular 
motors \cite{schliwa}. Not all molecular motors are, however, long-haul 
trucks. For example, muscle contraction associated with heartbeats are 
driven by a specific type of molecular motors which are specially 
designed for this purpose. So, it is not surprising that malfunctioning 
of the  molecular transport system can cause diseases- traffic disruption 
or traffic jam can bring an entire traffic system to a standstill. 

For obvious reasons, research on molecular motors has been a traditional 
area of research in molecular cell biology and biochemistry 
\cite{alberts}. However, in recent years, this area of research has 
attracted physicists \cite{fisher} as well as engineers \cite{mavroidis}. 
How do these motors work? Exploring the design and mechanisms of these 
motors from an {\it engineering perspective} \cite{mavroidis,barcohen} 
requires a investigation into their structure and dynamics using the 
fundamental principles of {\it physics} at the subcellular level. The 
insights gained from such fundamental research may find practical 
applications in designing and manufacturing artificial nano-motors. 
In contrast to man-made motors, the designs of natural nano-motors 
\cite{vale97,sakato,mermall} have evolved over billion of years. In 
this article, while discussing the design, mechanism and control of 
molecular motors, we'll make comparisons with their macroscopic 
counterparts to emphasize their common features as well as their 
differences. We do not report new data in this article. But, we carry 
out an in-depth qualitative investigation of the similarities and 
diferences between the intra-cellular and man-made macroscopic motors 
keeping in mind the mixed readership of this journal.

%%%%%%%%%%%%%%%%%%%%%%%%%%%%%%%%%%%%%%%%%%%%%%%%%%%%%%%%%%%%%
\subsection{Major components of the intracellular transport system} 
%%%%%%%%%%%%%%%%%%%%%%%%%%%%%%%%%%%%%%%%%%%%%%%%%%%%%%%%%%%%%

Just like the skeletons of human bodies, the cytoskeleton of an 
eukaryotic cell maintains its architecture. However, the cytoskeleton 
is not a rigid scaffold; it is a complex dynamic network that can 
change in response to external or internal signals. The cytoskeleton 
also serves as the network of tracks for the motors involved in 
intra-cellular transport processes. Moreover, the cytoskeleton plays 
important role in the motility of the cell as a whole.

%%%%%%%%%%%%%%%%%%%%%%%%%%%%%%%%%%%%%%%%%%%%%%%%%%%%%%%%%%%%%%
\begin{figure}[h]
\begin{center}
\includegraphics[angle=-90,width=0.5\columnwidth]{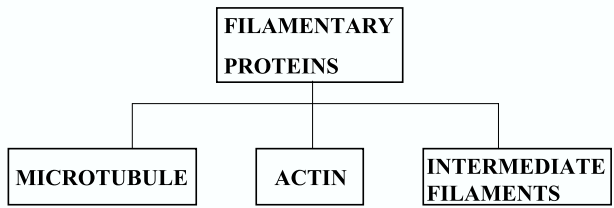}
\end{center}
\caption{The three classes of cytoskeletal filaments.
}
\label{fig-cytofila}
\end{figure}
%%%%%%%%%%%%%%%%%%%%%%%%%%%%%%%%%%%%%%%%%%%%%%%%%%%%%%%%%%%%%

%%%%%%%%%%%%%%%%%%%%%%%%%%%%%%%%%%%%%%%%%%%%%%%%%%%%%%%%%%%%%%
\begin{figure}[h]
\begin{center}
\includegraphics[angle=-90,width=0.5\columnwidth]{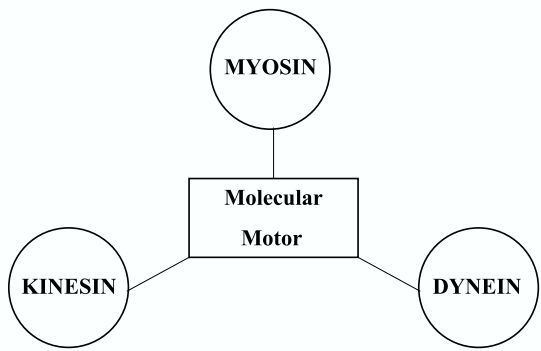}
\end{center}
\caption{The three superfamilies of cytoskeletal molecular motors.
}
\label{fig-motsupfam}
\end{figure}
%%%%%%%%%%%%%%%%%%%%%%%%%%%%%%%%%%%%%%%%%%%%%%%%%%%%%%%%%%%%%

The protein constituents of the cytoskeleton of eukaryotic cells 
can be broadly divided into the following three categories:  
(i) {\it Filamentous} proteins, (ii) {\it accessory} proteins, and 
(iii) {\it motor} proteins. The following three classes of 
filamentous proteins form the main scaffolding of the cytoskeleton 
(see fig.\ref{fig-cytofila}): (a) {\it actin}, (b) {\it microtubule}, 
and (c) {\it intermediate filaments}. 
The three superfamilies of cytoskeletal motor proteins \cite{howardbook} 
are (see fig.\ref{fig-motsupfam}) 
(i) {\it myosin} superfamily,
(ii) {\it kinesin} superfamily, and 
(iii) {\it dynein} superfamily.
Myosins either move on actin tracks or pull the actin filaments. In 
contrast, both kinesins and dyneins move on microtubules. 

Nucleic acids (DNA and RNA) also serve as tracks for another class 
of motors which are, more appropriately, also referred to as nucleic 
acid translocases \cite{michaelis07}. Just like cytoskeletal motors, 
these motors run on chemical fuel. But, instead of carrying cargo, 
these perform various other kinds of operations. For example, DNA 
helicase motors \cite{lohman} unzip the two strands of double-stranded 
DNA and use one of the two strands as the track for its directed 
movement. RNA polymerases \cite{wangrev,gelles} and ribosomes 
\cite{spirinbook} are mobile workshops which polymerize RNA and 
proteins, respectively, while moving on the corresponding specific 
tracks which also serve as the template for the synthesis.

%%%%%%%%%%%%%%%%%%%%%%%%%%%%%%%%%%%%%%%%%%%%%%%%%%%%%%%%%%%%%%5
\subsection{Processivity and Duty Ratio}
%%%%%%%%%%%%%%%%%%%%%%%%%%%%%%%%%%%%%%%%%%%%%%%%%%%%%%%%%%%%%%5

One of the key features of the dynamics of cytoskeletal motors is their 
ability to attach to and detach from the corresponding track. A motor 
is said to be attached to a track if at least one of its heads remains 
bound to one of the equispaced motor-binding sites on the corresponding 
track. Moreover, a motor can detach completely from its track.

One can define processivity in three different ways:\\
(i) Average number of {\it chemical cycles} before detachment from
the filament;\\
(ii) {\it attachment lifetime} of the motor to the filament;\\
(iii) {\it mean length} spanned by the motor on the filament in
a single run.\\
The first definition is intrinsic to the process arising from the
{\it mechano-chemical} coupling. But, it is extremely difficult to
measure experimentally. The other two quantities, on the other
hand, are accesible to experimental measurements.

During one cycle, suppose a motor spends an average time $\tau_on$
attached to the filament, and the remaining time $\tau_{off}$
detached from the filament. Clearly, the period during which it
exerts its {\it working} stroke is $\tau_{on}$ and its {\it recovery}
stroke takes time $\tau_{off}$. The {\it duty ratio}, $r$, is
defined as the fraction of the time that each head spends in its
attached phase, i.e.,
\begin{equation}
r = \tau_{on}/(\tau_{on} + \tau_{off}) 
\end{equation}
The typical duty ratios of kinesins and cytoplasmic dynein
are at least $1/2$ whereas that of conventional myosin can be
as small as $0.01$.

%%%%%%%%%%%%%%%%%%%%%%%%%%%%%%%%%%%%%%%%%%%%%%%%%%%%%%%5
\subsection{Design of filamentary tracks}
%%%%%%%%%%%%%%%%%%%%%%%%%%%%%%%%%%%%%%%%%%%%%%%%%%%%%%%5

\vspace{1cm}

%%%%%%%%%%%%%%%%%%%%%%%%%%%%%%%%%%%%%%%%%%%%%%%%%%%%%%%%%%%%%%
\begin{figure}[h]
\begin{center}
\includegraphics[angle=-90,width=0.35\columnwidth]{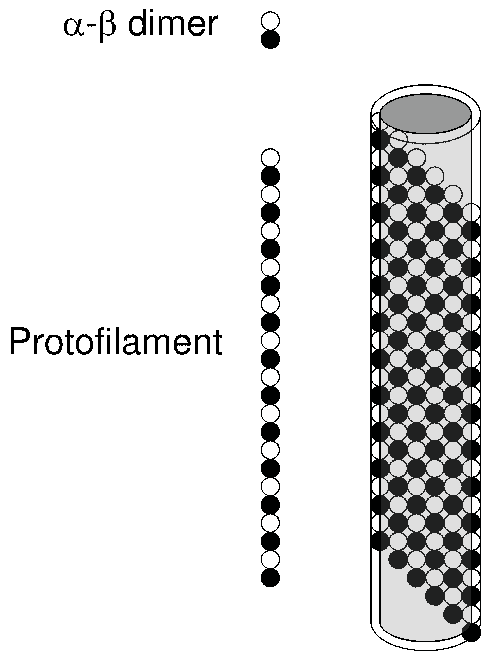}
\end{center}
\caption{A schematic presentation of a microtubule.
}
\label{fig-mtubule}
\end{figure}
%%%%%%%%%%%%%%%%%%%%%%%%%%%%%%%%%%%%%%%%%%%%%%%%%%%%%%%%%%%%%

Microtubules are cylindrical hollow tubes whose diameter is approximately 
20 nm (see fig.\ref{fig-mtubule}). The basic constituent of microtubules 
are globular proteins called tubulin. Hetero-dimers, formed by $\alpha$ 
and $\beta$ tubulins, assemble sequentially to form a protofilament. 
A sheet formed by the lateral organization of 13 such protofilaments 
then folds to form a microtubule. The length of each $\alpha-\beta$ dimer 
is 8 nm. Although the protofilaments are parallel to each other, there is 
a small offset of about 0.92 nm between the dimers of the neighbouring 
protofilaments. Thus, total offset accumulated over a single looping of 
the 13 protofilaments is $13 \times 0.92 \simeq 12 nm$ which is equal 
to the length of three $\alpha-\beta$ dimers joined sequentially. 
Therefore, the cylindrical shell of a microtubule can be viewed as 
{\it three} helices of monomers (see fig.\ref{fig-mtubule}). Moreover, 
the asymmetry of the hetero-dimeric building block and their parallel 
head-to-tail organization in all the protofilaments gives rise to the 
polar nature of the microtubules. The polarity of a microtubule is such 
an $\alpha$ tubulin is located at its - end and a $\beta$ tubulin is 
located at its + end. Majority of the kinesins are + end directed motors 
whereas most of the dyneins are - end directed motor proteins. Since 
there is only one binding site for a motor on each dimeric subunit of MT, 
the minimum step size for kinesins and dyneins is 8 nm.

Filamentous actin are polymers of globular actin monomers. Each 
actin filament can be viewed as a double-stranded, right handed 
helix where each strand is a single protofilament consisting of 
globular actin. The two constituent strands are half staggered 
with respect to each other such that the repeat period is 72 nm.
Majority of myosins are + end directed i.e., move towards the 
``barbed'' end of actin filaments \cite{howardbook}.

%%%%%%%%%%%%%%%%%%%%%%%%%%%%%%%%%%%%%%%%%%%%%%%%%%%%%%%5
\section{Design of motor proteins}
%%%%%%%%%%%%%%%%%%%%%%%%%%%%%%%%%%%%%%%%%%%%%%%%%%%%%%%5

A central location in these molecular motors is occupied by the 
ATPase site which binds to ATP. The motor protein acts like an 
enzyme and catalyzes the hydrolysis of the ATP, releasing the 
products ADP and phosphate. This enzymatic change causes small 
changes in the conformation of the protein surrounding the 
ATPase site which, in turn, propagates to farther regions and, 
ultimately, gets amplified into interdomain movements. It is these 
conformational changes that generate sufficiently large forces 
responsible for the unidirectional motion of the motor over long 
distances through repeated enzymatic cycles.

There are several architectural similarities between the three 
superfamilies of cytoskeleton-based motor proteins, namely, 
kinesin, dynein and myosin. All the motor proteins have at least 
two different functional domains:\\
(i) {\it head}: this domain contains a site for ATP hydrolysis 
and a binding site for attachment to a cytoskeletal filament; and \\ 
(ii) {\it stalk}. 
In addition, all kinesin and dyneins have a {\it tail} domain which 
binds with the cargo.

However, in spite of these general qualitative similarities, 
there are quantitative differences in their structural features 
and also striking differences in their biological functions. 
For example, the head domain of the kinesins is the smallest, that 
of myosins is of intermediate size whereas the head of dyneins is 
very large. The tail domain exhibits much more diversity than the 
head domain because of the necessity that the same motor should be 
able to recognize (and pick up) wide varieties of cargoes. Majority 
of the members of myosin and kinesin superfamilies are homodimers 
although hetero-dimeric kinesins have also been discovered. Some 
members of myosin and kinesin superfamilies are known to self-assemble 
into higher-order structures. The most well known among these 
higher-order structures is the myosin thick filaments in muscles.

According to the widely accepted nomenclature, myosins are classified 
into families bearing numerical (roman) suffixes (I, II, ..., etc.). 
Unlike conventional myosin-II of skeletal muscles, which has a very 
low duty ratio ($ \leq 0.05$), ``unconventional'' myosin-V and 
myosins-VI have quite high ($0.7-0.8$) duty ratios. Myosin-X has 
moderate duty ratio. Therefore, myosin-II are like ``rowers'' whereas 
myosin-V and myosin-VI are like ``porters''. Moreover, myosin-IX and 
myosin-VI are - end directed motors whereas all the other families of 
myosin are + end directed \cite{spudich}.

Kinesins are microtubule-based motor proteins. According to the latest 
standardized nomenclature of kinesins, the name of each family begins 
with the word ``kinesin'' followed by an arabic number ($1$, $2$, etc.). 
Kinesin-1, the double-headed ``conventional'' kinesin, are ``porters''. 
Dyneins are microtubule-based motor proteins. The molecular architecture 
of the dyneins is the most complex among the cytoskeletal motors. These 
motors consist of at least ten subunits in addition to the large motor 
domain.

%%%%%%%%%%%%%%%%%%%%%%%%%%%%%%%%%%%%%%%%%%%%%%%%%%%%%%%%%%%%%%
\begin{figure}[h]
\begin{center}
\includegraphics[angle=-90,width=0.5\columnwidth]{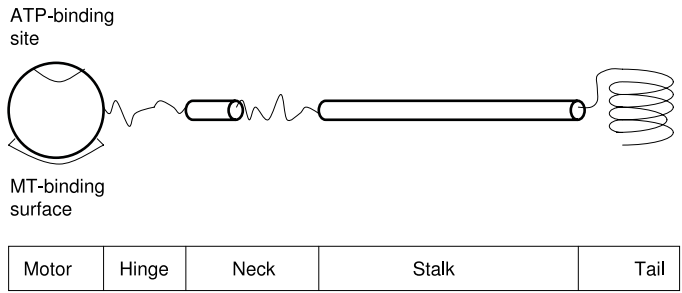}
\end{center}
\caption{A schematic representation of the organization of different 
domains in a typical kinesin motor (adapted from ref.\cite{ray06}).
}
\label{fig-motsupfam}
\end{figure}
%%%%%%%%%%%%%%%%%%%%%%%%%%%%%%%%%%%%%%%%%%%%%%%%%%%%%%%%%%%%%

%%%%%%%%%%%%%%%%%%%%%%%%%%%%%%%%%%%%%%%%%%%%%%%%%%%%%%%%%%%%%%%%5
\section{Fundamental questions}
%%%%%%%%%%%%%%%%%%%%%%%%%%%%%%%%%%%%%%%%%%%%%%%%%%%%%%%%%%%%%%%%5

In this section, we list some of the fundamental questions on the 
mechanism of operation of the motors as well as their regulation and 
control. We phrase the questions in such a way that these may appear 
to be directly relevant only for the cytoskeletal motors. But, these 
can be easily rephrased for the other types of motors including, for 
example, those which move on nucleic acid strands. These questions 
are as follows \cite{block}:\\
(i) {\bf Fuel}: In case of macroscopic motors, we normally begin with 
the question whether the engine of the motor runs on petrol, diesel, 
or electricity (or any other form of energy). The analogous question 
for molecular motors is: what is the {\it fuel} that supplies the 
(free-)energy input for the motor? The free energy released by the 
hydrolysis of ATP is usually the input for cytoskeletal motors. \\
(ii) {\it Engine, cycle and transmission}: The ATP-binding site on 
the motor, where ATP is hydrolyzed, can be identified as the engine
of the motor. What are the distinct states of the cyclic engine in
various stages of each cycle? Which step of the cycle is responsible
for the generation of force (or, torque)? How is the structural
(conformational) change, caused by this force (or torque), amplified
by the architecture of the motor? In other words, how does the
{\it trasmission} system of the motor work, i.e., what are the analogues
of the {\it clutch and gear} of automobiles? \\
(iii) {\it Track and traction}: What is the nature of the filamentous
track? Are they static or dynamic, i.e., do the lengths and/or
orientations of the tracks change with time? What is the traction
mechanism used by a motor head for staying on track, i.e., where is
the track-binding site located in the motor head and how does it
bind with the track? How is the traction controlled, i.e., how is the
affinity of the motor head for its track altered by its binding with
a fuel molecule, namely, ATP?\\
(iv) {\it Number of engines and coordination of their cycles}:
Recall that an internal combustion engine may have more than one cylinder 
where fuel is burnt and phased movements of the pistons of these 
cylinders gives rise to a practically smooth continuous motion of 
the common shaft to which they are connected. How many heads does each 
cytoskeletal motor possess? Are all the heads identical? If not, what 
functional advantages arise from such heterogeity? Are the cycles of 
the different engines of a motor coordinated in any manner and, if so, 
how is this coordination maintained? \\
(v) {\it Stroke and step}: Recall that, while crossing a shallow
stream by hopping on stones, the step size of a person is determined
not necessarily by the length of his/her legs but primarily by the
separation between the stones. A similar situation arises in the
processive movements of motor on their tracks and, therefore, one
has to distinguish between size of a {\it stroke} and that of a
{\it step} \cite{howardbp06}. The separation between the two successive
binding sites on the track is the smallest possible step size of the
motor. On the other hand, a stroke is a conformational change of the
motor bound to the track and it takes the motor closer to its next
prospective binding site on the track. In general, the stroke size
need not be equal to the step size. If the motor covers only a fraction
of the distance to the next binding site by the stroke, how does it
manage to cover the remaining distance?  Can the same motor adopt
different step sizes under different circumstances? What features of
the track or/and the motor determines the step size? \\
(vi) {\it Directionality and processivity}: What determines the 
direction of movement, i.e., why are some motors $+$-end directed
whereas the others are $-$-end directed? Can a motor reverse its 
direction of motion (a) spontaneously, or (b) under an opposing 
(load) force?  Do the motors possess reverse gears and is it possible 
to reverse the direction of their movement by utilizing the reverse 
gear mechanism? What is the minimal change (e.g., mutation) required 
to reverse the direction of motion of a motor? What is the mechanism 
that decides the {\it processivity} (or the lack of processivity) of 
a motor? The directionality and processivity of cytoskeletal motors 
involve coordination of essentially three cycles \cite{milliganbp06}:
(a) ATP hydrolysis cycle, (b) the motor head-track binding cycle
(periodic attachments and detachments of each motor head from the track),
and (c) conformational cycle of the motor.  \\
(vii) {\it Stepping pattern}:
Does the motor move like an ``inchworm'' or does the stepping appear
more like a ``hand-over-hand'' mechanism? Moreover, two types of
hand-over-hand mechanism are possible: symmetric and asymmetric.
In the symmetric pattern, the two heads exchange positions, but the
three-dimensional structure of the molecule is preserved at all equivalent
positions in the cycle. In contrast, in the asymmetric pattern, the two
heads exchange position, but alternate steps differ in some way, e.g.,
what happens in ``limping'' which involves alternate faster and slower
stepping phases. Can a motor switch from one track to a neighbouring
track and, if so, how does it achieve that? What prevents a motor from
changing lane on a multi-lane track? \\
(viii) {\it Speed and efficiency}: Is the average speed of a processive
motor determined by the track or the motor or fuel or some external 
control mechanism? Recall that the average speed of a car on a highway 
in sparse traffic can be decided either by the smoothness of the highway,
or by the model of the car (whether it is a Ferrari or a heavy truck), 
or by the quality of the fuel. Similarly, how does the molecular 
constitution of the track and the nature of the motor-track interaction 
affect the speed of the motor? Can an external force applied to a motor 
in the forward direction speed it up?  Is the mechano-chemical coupling 
{\it tight} or {\it loose}? If hydrolysis of ATP provides the input 
free energy, then, how many steps does the motor take for every molecule 
of ATP hydrolyzed, or, equivalently, how many ATP molecules are consumed 
per step of the motor ? How does the speed of the motor depend on the 
opposing ``load'' force? What is the maximum speed it can attain? What 
is the stalling load force? What is the most appropriate definition of 
efficiency of the motor and how to estimate that efficiency? \\
(ix) {\it Regulation and control}: How is the operation of the motor
{\it regulated}? For example, how is the motor switched on and off?
Recall that the speed of a car can also be regulated by imposing the
some speed limit or by traffic signals. Are there molecular {\it
signals} that control the motor's movement on its track and how?
How do motors get back to their starting points of the processive run 
after delivering their cargo?\\
(x) {\it Cargo}: What kinds of cargoes can be hauled by the motor?
How does the motor pick up its cargo and how does it drop it at the
target location? \\
(xi) {\it Motor-motor interactions}: How do different types of motors 
interact while moving on the same track carrying their cargo? How do 
different classes of motors, which move on different types of tracks, 
coordinate their functions and even transfer or exchange their cargoes? \\

%%%%%%%%%%%%%%%%%%%%%%%%%%%%%%%%%%%%%%%%%%%%%%%%%%%%%%%%%%%%%%%%%%%%
\section{Macro- and nano-motors: a comparison}
%%%%%%%%%%%%%%%%%%%%%%%%%%%%%%%%%%%%%%%%%%%%%%%%%%%%%%%%%%%%%%%%%%%%

Several superficial qualitative analogies between the components of
engines of macroscopic motorized vehicles (e.g., internal combustion
engine) and those of cytoskeletal nano-motors are listed in table
\ref{table-compare}. Moreover, qualitative similarities and quantitative
differences between the processes underlying their operational
mechanisms are also listed for the purpose of comparison.

%%%%%%%%%%%%%%%%%%%%%%%%%%%%%%%%%%%%%%%%%%%%%%%%%%%%%%%%%%%%%%%%%%%%%
\begin{table}
\begin{tabular}{|c|c|} \hline
Internal combustion engine  & Coventional 2-headed kinesin \\\hline
4 cylinders & 2 heads \\ \hline
Combustion chamber & ATP-binding site \\ \hline
Firing the fuel & ATP hydrolysis \\ \hline
4 strokes of the Otto cycle in each cylinder & 4 chemical states of the mechano-chemical cycle of each head \\ \hline
Common crank shaft connected to 4 cylinders & Stalk connected to 2 heads \\ \hline
Typical efficiency $\simeq 20 \%$ & Typical efficiency $\simeq 50 \%$ \\ \hline
\end{tabular}
\caption{Comparison of 4-cylinder internal combustion engine of macroscopic engines and 2-headed conventional kinesin nano-motor.
 }
\label{table-compare}
\end{table}
%%%%%%%%%%%%%%%%%%%%%%%%%%%%%%%%%%%%%%%%%%%%%%%%%%%%%%%%%%%

Biomolecular motors operate in a domain where the appropriate units of 
length, time, force and energy are {\it nano-meter}, {\it milli-second}, 
{\it pico-Newton} and $k_B T$, respectively ($k_B$ being the Boltzmann 
constant and $T$ is the absolute temperature). From table 
\ref{table-compare}, naively, one may think that the differences in 
the mechanisms of the molecular and macroscopic machines is merely a 
matter of two different scales (of size, time, force, energy, etc.). 
But, that is not true. 

Since the masses of the molecular motors are extremely small, they 
are subjected to two dominating forces which are quite small for 
all the man-made macroscopic motors. Since the inertial forces are 
small compared to the viscous force, the dynamics of molecular motors 
is dominated by hydrodynamics at low Reynold's number. Moreover, the 
nano-motors are bombarded from all sides by the randomly moving water 
molecules. Because of these bombardments, the molecular motors 
experience an additional random force which leads to ``noisy'' 
trajectories of the motors.  Furthermore, the active processes 
\cite{julicher06}, in which these motors are involved, cannot be 
described by equilibrium statistical mechanics. Finally, it is worth 
pointing out that, unlike man-made motors, these natural nano-motors 
are capable of transducing chemical energy directly into mechanical work.

%%%%%%%%%%%%%%%%%%%%%%%%%%%%%%%%%%%%%%%%%%%%%%%%%%%%%%%%%%%%%%
\begin{figure}[h]
\begin{center}
\includegraphics[angle=-90,width=0.5\columnwidth]{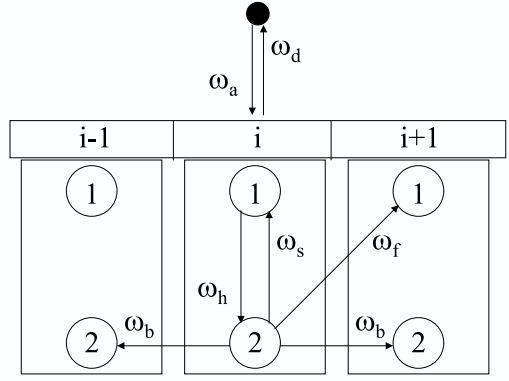}
\end{center}
\caption{A schematic description of the model reported in ref.\cite{nosc} 
for a single-headed kinesin motor KIF1A. The symbols $...,i-1,i,i+1,...$ 
label the motor-binding sites on the microtubule track. The allowed 
transitions are indicated by the arrows together with the corresponding 
rate constants, i.e., transition probabilities per unit time. Note 
that $\omega_{a}$ accounts for the possibility of attachment of 
motors to the microtubule track whereas $\omega_{d}$ accounts for 
the reverse process, i.e., detachment of bound motors (from 
ref.\cite{greulich}).
}
\label{fig-kifmod}
\end{figure}
%%%%%%%%%%%%%%%%%%%%%%%%%%%%%%%%%%%%%%%%%%%%%%%%%%%%%%%%%%%%%

%%%%%%%%%%%%%%%%%%%%%%%%%%%%%%%%%%%%%%%%%%%%%%%%%%%%%%%%%%%%
\section{Modeling and simulation at different levels}
%%%%%%%%%%%%%%%%%%%%%%%%%%%%%%%%%%%%%%%%%%%%%%%%%%%%%%%%%%%%

Modeling molecular motors is a problem of mechano-chemistry; one has to 
capture the interplay of mechanical movements and chemical reactions. 
Such models can be developed at different {\it levels} of molecular 
details \cite{wang}. Even at a given level, the dynamics of the system 
can be formulated using different types of formalisms or updating rules 
for computer simulations.

(i) {\it molecular} level: In principle, one can write down the 
Newton's equations for all the atomic constituents of the motor and 
those of the surrounding aqueous medium. But, in practice, no molecular 
dynamics simulation of molecular motors is possible in forseeable 
future using these equations, because the duration of a single processive 
run of a cytoskeletal motor on its track is several orders of magnitude 
longer than the longest molecular dynamics simulation possible at present.

(ii) {\it Brownian} level: At this level one can treat the motor as 
a Brownian particle that moves in an energy landscape determined by 
the chemical reaction.  For simplicity, we assume the motion to be
restricted in one-dimensional space. The effects of the chemical
reactions (e.g., ATP hydrolysis, etc.) are incorporated by asigning
``internal'' or ''chemical states'' to the motor. Therefore, in the
overdamped regime, the dynamics of the center of mass of the motor
$x$ obeys the Langevin equation 
\begin{equation}
0 = - \gamma \frac{dx}{dt} - \frac{dV_{\mu}(x)}{dx} + F_{ext} + \xi(t) 
\end{equation}
where $V_{\mu}(x)$ is the potential experienced by the motor at the
position $x(t)$ when it is in the ``chemical'' state $\mu$. Moreover,
as usual, $F_{ext}$ is the externally applied mechanical force (the 
sign of the term will be negative in case of a load force) and
$\xi(t)$ is the random Brownian force. The potential $V_{\mu}(x)$
evolves with time because of the chemical transitions. The chemical
state evolves following the discrete master equation
\begin{eqnarray}
\frac{\partial P_{\mu}(x,t)}{\partial t} = \sum_{\mu'} P_{\mu'}(x,t) W_{\mu'\to \mu}(x) - \sum_{\mu'} P_{\mu}(x,t) W_{\mu \to \mu'}(x)
\end{eqnarray}

In order to formulate the equivalent Fokker-Planck equations (more
appropriately, a hybrid of Fokker-Planck and master equations), we
define the probability $P_{\mu}(x,t)$ that at time $t$ the center of
mass of the motor is located at $x$ while it is in the discrete
(internal) ``chemical state'' $\mu$. The equation of motion governing
the time evolution of $P_{\mu}(x,t)$ is a combination of a Fokker-Planck
equation and a master equation; the Fokker-Planck part describes the
dynamics in continuous space while the Master equation accounts for
the dynamics of transitions between discrete chemical states.

\begin{eqnarray}
\frac{\partial P_{\mu}(x,t)}{\partial t} &=& \frac{1}{\eta} \frac{\partial}{\partial x}\biggl[\{V'_{\mu}(x) - F\} P_{\mu}(x,t)\biggr] + \biggl(\frac{k_BT}{\eta}\biggr) \frac{\partial^2 P_{\mu}(x,t)}{\partial x^2} \nonumber \\
&+& \sum_{\mu'} P_{\mu'}(x,t) W_{\mu'\to \mu}(x) - \sum_{\mu'} P_{\mu}(x,t) W_{\mu\to \mu'}(x)
\end{eqnarray}

The Brownian level modeling has served well in elucidating the generic 
principles involved in the mechanisms of their directed transport 
\cite{julicher,reimann}. But, one of the difficulties of using this 
formalism for any specific member of a particular superfamily of motor 
proteins is that the potential $V_{\mu}(x)$ experienced by the motor 
is not available (unless derived from a more microscopic model for the 
combined motor-track system).

(iii) {\it Fully discretized chemical kinetic} level: 

The kinetic rate equation formalisms are at a level higher than the
Brownian level. In this approach one assumes a set of {\it discrete}
states (i.e., discrete positions and discrete velocities) of the
motor and the transitions between these states are given by appropriately
chosen rate constants \cite{fisher}. Some of the rate constants can 
depend on force
and the form of the force-dependence is postulated on some physical
grounds. This approach is based on the assumption that the phase
space of the entire system, which is analogoues to a landscape, can
be conceptually divided into a finite number of distinct regions each
of which is like a deep valley in the landscape. A purely rate equation
approach with discrete states is a reasonable approximation provided
the barriers separating the valleys are sufficiently high. In principle,
the rate constants can be derived from a more microscopic description
like, for example, theories at the Brownian level. Alternatively, the 
phenomenological rate constants can be extracted from one set of 
empirical data and, then, can be used in all other situations for the 
same motor system.

%%%%%%%%%%%%%%%%%%%%%%%%%%%%%%%%%%%%%%%%%%%%%%%%%%%%%%%%%%%%%%%%%%%%%
\begin{table}
\begin{tabular}{|c|c|c|c|c|} \hline
ATP (mM)  & $\omega_h ~(s^{-1})$ & $v$ (nm/ms) & $D/v$ (nm) & $\tau$ (s) \\\hline
$\infty$ & 250 & 0.201 & 184.8 & 7.22 \\ \hline
0.9 & 200 & 0.176 & 179.1 & 6.94 \\ \hline 
0.3375 & 150 & 0.153 & 188.2 & 6.98 \\ \hline 
0.15 & 100 & 0.124 & 178.7 & 6.62 \\ \hline
\end{tabular}
\caption{Transport properties of single-headed kinesin, at four different 
concentrations of ATP molecules, obtained from computer simulation of 
the model shown in fig.\ref{fig-kifmod} (from ref.\cite{greulich}).
 }
\label{tab-kifres}
\end{table}
%%%%%%%%%%%%%%%%%%%%%%%%%%%%%%%%%%%%%%%%%%%%%%%%%%%%%%%%%%%

%%%%%%%%%%%%%%%%%%%%%%%%%%%%%%%%%%%%%%%%%%%%%%%%%%%%%%%%%%%%%%%
\section{Kinesin motor on MT track}
%%%%%%%%%%%%%%%%%%%%%%%%%%%%%%%%%%%%%%%%%%%%%%%%%%%%%%%%%%%%%%%

It is now well established that the processivity of the double-headed 
conventional kinesin motor is, at least partly, because of the well 
coordinated out-of-phase ATPase cycles of the two heads which enables 
one of the heads to remain bound to the MT track while the other steps 
forward. However, an altogether different mechanism, based on Brownian 
ratchet concept \cite{julicher,reimann}, had to be invoked to explain 
the experimentally observed processivity of single-headed kinesin. 
The model is shown shematically in fig.\ref{fig-kifmod}. In the two 
``chemical'' states labelled by 1 and 2 the motor head is bound, 
respectively, strongly and weakly to the MT track. This is a multi-step 
stochastic chemical kinetic model where both positions and chemical 
states of a motor are discrete. The detailed justification of this 
two-state model and the physical interpretation of the allowed 
transitions are given in ref.\cite{nosc}. 

Carrying out computer simulations of this model, we computed some of 
the transport properties of the single-headed kinesin molecule; the 
results are listed in table \ref{tab-kifres}. In this table $v$ is 
the average speed, $D$ is the diffusion constant and $\tau$ is the 
average run time of the motors. These predicted values are in good 
quantitative agreement with the corresponding results obtained from 
{\it in-vitro} single molecule experiments. 

\vspace{2cm} 

%%%%%%%%%%%%%%%%%%%%%%%%%%%%%%%%%%%%%%%%%%%%%%%%%%%%%%%%%%%%%%
\begin{figure}[h]
\begin{center}
\includegraphics[angle=-90,width=0.35\columnwidth]{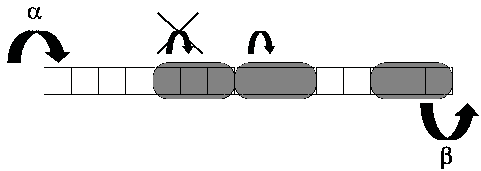}
\end{center}
\caption{A schematic description of the collective movement of 
interacting ribosomes on a single mRNA track. 
}
\label{fig-moltraf}
\end{figure}
%%%%%%%%%%%%%%%%%%%%%%%%%%%%%%%%%%%%%%%%%%%%%%%%%%%%%%%%%%%%%

%%%%%%%%%%%%%%%%%%%%%%%%%%%%%%%%%%%%%%%%%%%%%%%%%%%%%%%%%%%%%%%
\section{Ribosome traffic on mRNA track: fully discretized chemical kinetic model}
%%%%%%%%%%%%%%%%%%%%%%%%%%%%%%%%%%%%%%%%%%%%%%%%%%%%%%%%%%%%%%%

%%%%%%%%%%%%%%%%%%%%%%%%%%%%%%%%%%%%%%%%%%%%%%%%%%%%%%%%%%%%%%
\begin{figure}[h]
\begin{center}
\includegraphics[angle=-90,width=0.35\columnwidth]{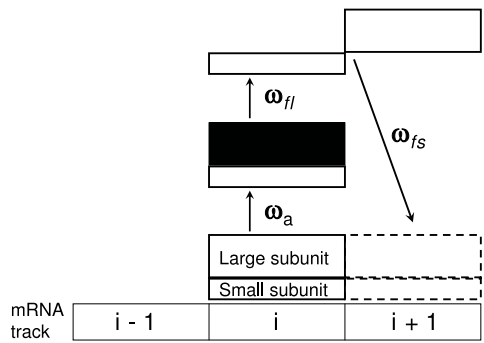}
\end{center}
\caption{A simple 3 state model of a ribosome during the elongation 
stage of protein synthesis.
}
\label{fig-ribomodel}
\end{figure}
%%%%%%%%%%%%%%%%%%%%%%%%%%%%%%%%%%%%%%%%%%%%%%%%%%%%%%%%%%%%%

%%%%%%%%%%%%%%%%%%%%%%%%%%%%%%%%%%%%%%%%%%%%%%%%%%%%%%%%%%%%%%
\begin{figure}[h]
\begin{center}
\includegraphics[width=0.35\columnwidth]{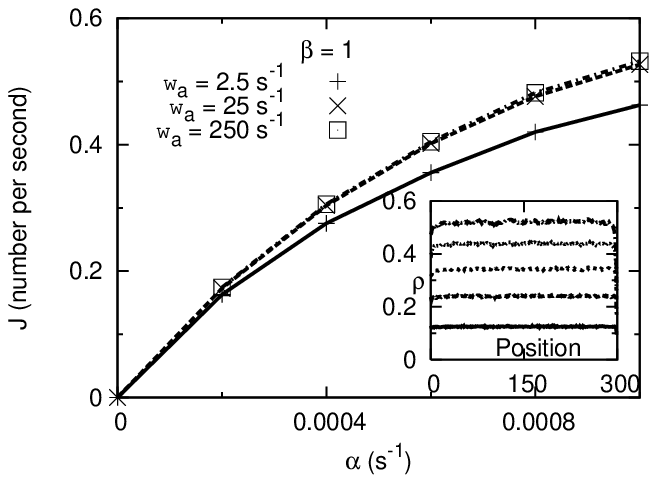}
\includegraphics[width=0.35\columnwidth]{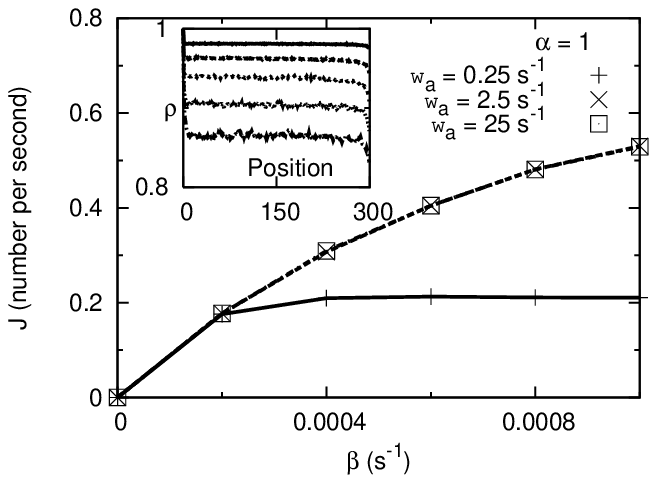}
\end{center}
\caption{Total rate of protein synthesis by ribosomes under open 
boundary conditions plotted against $\alpha$ in (a) and $\beta$ in (b) 
for three values of $\omega_{a}$. The discrete data points were 
obtained by doing computer simulations, and the curves are merely
guides to the eye. The average density profiles are plotted in the
insets. In the inset of (a) the lowermost density profile corresponds
to $\alpha=0.0002$, and the topmost one corresponds to $\alpha=0.001$;
$\alpha$ varies from one profile to the next in steps of $0.0002$.
In the inset of (b) the topmost density profile corresponds to
$\beta=0.0002$, and the lowermost one corresponds to $\beta=0.001$;
$\beta$ varies from one profile to the next in steps of $0.0002$. 
The other parameters are $\omega_{f\ell} = 1.8 ~s^{-1}$ and 
$\omega_{fs} = 10 ~s^{-1}$. 
(from ref.\cite{basuajp}).
}
\label{fig-ribores}
\end{figure}
%%%%%%%%%%%%%%%%%%%%%%%%%%%%%%%%%%%%%%%%%%%%%%%%%%%%%%%%%%%%%

A protein is a bio-polymer, each characterized by a specific sequence 
of its monomeric subunits, called amino acids. Since these subunits 
are covalebtly linked by peptide bonds, proteins are also referred to 
as polypeptides. The polymerization of a polypeptide is carried out 
by a macromolecular complex, called ribosome \cite{spirinbook}, 
following the instructions encoded in the sequence of nucleotides in 
the template mRNA on which the ribosome moves as a motor. We assume 
that the initiation and termination of protein synthesis by each 
ribosome takes place with the rates $\alpha$ and $\beta$ respectively; 
these rates are captured by imposing open boundary cnditions (OBC) on 
the system as shown in fig.\ref{fig-moltraf}. Since a ribosome is much 
larger than a codon (a triplet of nucleotides) and we assume that each 
ribosome can simultaneously  cover $r$ lattice sites where each lattice 
site corresponds to a single codon. However, the position of a ribosome 
is denoted by that of the leftmost site it covers. Moreover, at each 
step a ribosome moves forward by one codon, i.e., a single lattice 
site. In order to capture the hard core steric interactions between 
the ribosomes, we impose the condition that no codon can be covered 
simultaneously by more than one ribosome.

For the purpose here, it is sufficient to note that (i) each ribosome 
consists of two subunits, called {\it large} and {\it small} subunits, 
respectively; (ii) the three major steps in the mechano-chemical cycle 
of a ribosome during the elongation of the polypeptide are as follows 
\cite{basuajp}:\\
(a) {\it arrival} of the correct amino-acid subunit in association 
with an adaptor molecule which, with the help of the ribosome, decodes 
the genetic message from the template track;\\
(b) following the formation of the peptide bond between the growing 
polypeptide and the newly selected amino acid, the larger subunit steps 
ahead by a codon on the mRNA track;\\ 
(c) stripped of its amino-acid subunit, which is now bonded to the 
growing polypeptide, the adapter molecule makes its final exit from 
the ribosome and the smaller subunit also steps forward to the next 
codon.\\ 
These three steps of the mechano-chemical cycle are shown schematically 
in fig.\ref{fig-ribomodel} together with the corresponding rate constants.
We'll not go into the finer details of the processes involved in 
polypeptide synthesis \cite{basupre}. 

We define the flux $J$ as the average number of ribosomes crossing 
the stop codon per unit time. The flux of the ribosomes gives the 
total rate of synthesis of the polypeptides. Although it is possible 
to compute this quantity for any given codon sequence on the mRNA 
template \cite{basupre}, we present here the data for the simple case 
of a homogeneous sequence \cite{basuajp}. 

The flux of the ribosomes, 
i.e., the total rate of protein synthesis, obtained from computer 
simulations of the model defined by the figs. \ref{fig-moltraf} and 
\ref{fig-ribomodel} are plotted in fig.\ref{fig-ribores}(a) and 
fig.\ref{fig-ribores}(b) as functions of $\alpha$ and $\beta$, 
respectively. The average density profiles observed for several 
values of $\alpha$ and $\beta$ are also shown in the insets of 
figs.~\ref{fig-ribores}(a) and (b). For $\alpha < \beta = 1$, the 
flux increases, and gradually saturates as $\alpha$ increases 
because higher $\alpha$ corresponds to a higher rate of initiation 
of protein synthesis. This increase of flux with increasing $\alpha$ 
is also consistent with the corresponding higher average density 
profile shown in the inset of fig.~\ref{fig-ribores}(a). 
For $\beta < \alpha = 1$, the increase, and gradual saturation, of 
flux with increasing $\beta$ is caused by the weakening of the 
bottleneck at the stop codon. This trend of variation of flux with
$\beta$ is also consistent with the gradual lowering of the average
density profile with increasing $\beta$ (see the inset of
fig.~\ref{fig-ribores}(b). Similar results have been reported recently 
\cite{tripathi} also for the traffic of RNA polymerase motors on 
DNA tracks during the synthesis of RNA.

%%%%%%%%%%%%%%%%%%%%%%%%%%%%%%%%%%%%%%%%%%%%%%%%%%%%%%%%%%%%%%%
\section{Summary and conclusion}
%%%%%%%%%%%%%%%%%%%%%%%%%%%%%%%%%%%%%%%%%%%%%%%%%%%%%%%%%%%%%%%

In this article we have illustrated the use of models and computer 
simulations for studying the mechanisms of molecular motors by 
applying the technique to two particular cases. A list of some 
common intracellular molecular motors and their filamentary tracks 
is given in table \ref{tab-motlist}; however, this list is far from 
complete.

%%%%%%%%%%%%%%%%%%%%%%%%%%%%%%%%%%%%%%%%%%%%%%%%%%%%%%%%%%%%%%%%%%%%%
\begin{table}
\begin{tabular}{|c|c|} \hline
Motor  & Track \\\hline
Dynein & Microtubule \\ \hline
Kinesin & Microtubule \\ \hline
Myosinsin & Actin filament \\ \hline
DNA helicase & ssDNA \\ \hline
RNA helicase & RNA \\ \hline
DNA polymerase & ssDNA \\ \hline
RNA polymerase & ssDNA \\ \hline
Ribosome & mRNA \\ \hline
\end{tabular}
\caption{Some common intracellular molecular motors and the corresponding tracks.  }
\label{tab-motlist}
\end{table}
%%%%%%%%%%%%%%%%%%%%%%%%%%%%%%%%%%%%%%%%%%%%%%%%%%%%%%%%%%%

How are the motors regulated? The experimental data strongly indicate 
that, perhaps, there is no universal or generic mechanism for 
regulating the operation of molecular motors. It is widely believed 
that, because of the higher architectural complexity and association 
with several accessory proteins, dyneins may be regulated in several 
ways which are not possible in the simpler cases of kinesin and 
myosin motors \cite{mallik}. The mechano-chemical pathways of motors 
are related to their function. The speed of operation of a motor can 
be controlled either by varying the concentration of the fuel or 
ligands or by applying external force which alters the rate of the 
chemical reactions \cite{busta}. 

It is amusing to note that molecular motors have no wheel. The 
cytoskeletal motors we have considered in this article are, at least 
structurally, more like ``porters'' walking along a track carrying load 
on their heads. In fact, there are some related non-processive cytoskeletal 
motors which work collectively in a manner that has strong similarity with 
``rowers'' of boats. In addition to the ``linear'' motors, 
which move on filaments, cells also use rotary motors. In fact, to our 
knowledge, the smallest rotary motor is the ATP synthase \cite{oster} 
which is used by every animal cell to synthesize ATP, the most common fuel 
for intracellular machineries. Another rotary motor, somewhat larger than 
ATP synthase, is the bacterial flagellar motor \cite{berg} which rotates 
the flagellum that is responsible for bacterial locomotion in aqueous media.
We hope to report exciting new developments in computer simulations of 
some of these motors in near future. 

Efforts are being made to exploit the lessons learnt from the studies 
of natural nano-motors to design and manufacture artificial nano-motors 
\cite{heuvel}. Two different strategies are being following in such 
bottom-up approaches. In one of these integrates the components of 
the natural motors into an artificial scaffold so as to get the required 
performance of the machine. In the alternative approach, one synthesizes 
a fully artificial motor molecule whose design mimics the design of its 
natural counterpart. However, several practical hurdles remain on the 
path of commercialization of nano-machines. Neverthless, nano-robotics 
\cite{mavroidis} may no longer be a distant dream.

%%%%%%%%%%%%%%%%%%%%%%%%%%%%%%%%%%%%%%%%%%%%%%%%%%%%%%%%%%%%%%%
{\bf Acknowledgements}: I thank A. Basu, A. Garai, M. Gopalakrishnan, 
B. S. Govindan, P. Greulich, K. Nishinari, Y. Okada, T. V. Ramakrishnan, 
A. Schadschneider, T. Tripathi and J.S. Wang for enjoyable and fruitful 
collaborations. It is my great pleasure to thank B. Lindner, R. Mallik 
and K. Ray for critical reading of the manuscript and for useful 
suggestions. I also thank  J. Howard and F. J\"ulicher for useful 
discussions on cytoskeletal motors. 
This work is supported, in part, by Council of Scintific and Industrial 
Research (India) and the Visitors Program of the Max-Planck Institute 
for Physics of Complex Systems, Dresden (Germany).
%%%%%%%%%%%%%%%%%%%%%%%%%%%%%%%%%%%%%%%%%%%%%%%%%%%%%%%%%%%%%%%

%%%%%%%%%%%%%%%%%%%%%%%%%%%%%%%%%%%%%%%%%%%%%%%%%%%%%%%%%%%%%%%


\begin{thebibliography}{99}
%%%%%%%%%%%%%%%%%%%%%%%%%%%%%%%%%%%%%%%%%%%%%%%%%%%%%%%%%%%%%%%

\bibitem{schliwa} M. Schliwa, (ed.) {\it Molecular Motors}, (Wiley-VCH, 2003).
\bibitem{alberts} B. Alberts, {\it The cell as a collection of 
protein machines: preparing the next generation of molecular 
biologists}, Cell {\bf 92}(3), 291-294 (1998).
\bibitem{fisher} A.B. Kolomeisky and M.E. Fisher, {\it Molecular motors: 
a theorist's perspective}, Annu. Rev. Phys. Chem. {\bf 58}, 675-695 
(2007). 
\bibitem{mavroidis} C. Mavroidis, A. Dubey and M.L. Yarmush, {\it 
Molecular Machines}, in: {\it Annual Rev. Biomed. Engg.}, {\bf 6}, 
363-395 (2004). 
\bibitem{barcohen} Y. Bar-Cohen, ed. {\it Biomimetics: biologically 
inspired technologies} (Taylor and Francies, 2005). 
\bibitem{vale97} R.D. Vale and R.J. Fletterick, {\it The design plan 
of kinesin motors}, in: {\it Annu. Rev.  Cell Dev. Biol.}, {\bf 13}, 
745-777 (1997).
\bibitem{sakato} M. Sakato and S.M. King, {\it Design and regulation 
of the AAA+ microtubule motor dynein}, J. Struc. Biol. {\bf 146}(1-2), 
58-71 (2004).
\bibitem{mermall} V. Mermall, P.L. Post and M.S. Mooseker, {\it 
Unconventional myosins in cell movement, membrane traffic, and signal 
transduction}, Science {\bf 279}(5350), 527-533 (1998). 
\bibitem{howardbook} J. Howard, {\it Mechanics of motor proteins and the
cytoskeleton}, (Sinauer Associates, 2001).
\bibitem{michaelis07} K.P. Hopfner and J. Michaelis, {\it Mechanisms 
of nucleic acid translocases: lessons from structural biology and 
single-molecule biophysics}, Curr. Op. in Str. Biol. {bf 17}(1), 87-95 
(2007). 
\bibitem{lohman} T.M. Lohman, K. Thorn and R.D. Vale, {\it Staying on 
track: common features of DNA helicases and microtubule motors}, Cell 
{\bf 93}(1), 9-12 (1998).
\bibitem{wangrev} L. Bai, T.J. Santangelo and M.D. Wang, {\it 
Single-molecule analysis of RNA polymerase transcription}, Annu. 
Rev. Biophys. Biomol. Str. {\bf 35}, 343-360 (2006).
\bibitem{gelles} J. Gelles and R. Landick, {it RNA polymerase as a 
molecular motor}, Cell {\bf 93}(1), 13-16 (1998).
\bibitem{spirinbook} A. S. Spirin, {\it Ribosomes}, (Springer, 2000).
\bibitem{ray06} K. Ray, {\it How kinesins walk, assemble and transport: 
a birds-eye-view of some unresolved questions}, Physica A {\bf 372}(1), 
52-64 (2006). 
\bibitem{spudich} R.S. Rock, T.J. Purcell and J.A. Spudich, {\it 
Mechanics of unconventional myosins}, in: {\it The Enzymes: energy 
coupling and molecular motors}, eds. D.D. Hackney and F. Tamanoi 
(Elsevier, 2004).
\bibitem{block} S.M. Block, {it Kinesin motor mechanics: binding, stepping, 
tracking, gating and limping}, Biophys. J. {\bf 92}(9), 2986-2995 (2007). 
\bibitem{howardbp06} J. Howard, in: proceedings of 2006 Biophysical 
Society Discussions on ``Molecular Motors: Point Counterpoint'', October 
19-22, 2006 (available online at 
http://www.biophysics.org/discussion/2006/book1.pdf).
\bibitem{milliganbp06} R.A. Milligan and C. Yoshioka, {\it Motor 
directionality in the kinesins},  in: proceedings of 2006 
Biophysical Society Discussions on ``Molecular Motors: Point Counterpoint'', 
October 19-22, 2006 (available online at
http://www.biophysics.org/discussion/2006/book1.pdf).
\bibitem{julicher06} F. J\"ulicher, {\it Statistical physics of active 
processes in cells}, Physica A {\bf 369}(1), 185-200 (2006).
\bibitem{wang} H. Wang and T.C. Elston, {\it Mathematical and computational 
methods for studying energy transduction in protein motors}, J. Stat. 
Phys, {\bf 128}(1), 35-76 (2007). 
\bibitem{julicher} F. J\"ulicher, A. Ajdari and J. Prost, {\it Moceling 
molecular motors}, Rev. Mod. Phys.  {\bf 69}(4), 1269-1281 (1997).
\bibitem{reimann} P. Reimann, {\it Brownian motors: noisy transport far 
from equilibrium}, Phys. Rep.{\bf 361}(2-4), 57-265 (2002). 
\bibitem{nosc} K. Nishinari, Y. Okada, A. Schadscneider and D. Chowdhury, 
{\it Intracellular transport of single-headed molecular motors KIF1A}, 
Phys. Rev. Lett. {\bf 95}(11), 118101 (2005). 
\bibitem{greulich} P. Greulich, A. Garai, K. Nishinari, A. Scahschneider 
and D. Chowdhury, {\it Intracellular transport by single-headed kinesin 
KIF1A: effects of single-motor mechanochmistry and steric interactions}, 
Phys. Rev. E {\bf 75}(4), 041905 (2007). 
\bibitem{basupre} A. Basu and D. Chowdhury, {\it Traffic of interacting 
ribosomes: effects of single-machine mechanochemistry on protein synthesis}, 
Phys. Rev. E {\bf 75}(2), 021902 (2007). 
\bibitem{basuajp} A. Basu and D. Chowdhury, {\it Modeling protein 
synthesis from a physicist's perspective: a toy model}, Am. J. Phys. 
{\bf 75}, 931-937 (2007). 
\bibitem{tripathi} T. Tripathi and D. Chowdhury, {\it RNA polymerase 
motors on DNA track: effects of traffic congestion and intrinsic noise 
on protein synthesis}, arXiv:0708:1067 (2007). 
\bibitem{mallik} R. Mallik and S.P. Gross, {\it Molecular motors: 
strategies to get along}, Curr. Biol. {\bf 14}(22), R971-R982 (2004). 
\bibitem{busta} C. Bustamante, Y.R. Chemla, N.R. Forde and D. Izhaky, 
{\it Mechanical processes in biochem.}, Annu. Rev. Biochem. {\bf 73}, 
705-748 (2004). 
\bibitem{oster} G. Oster and H. Wang, {\it Reverse engineering a protein: 
the mechanochemistry of ATP synthase}, Biochim. et Biophys. Acta 
(Bioenergetics) {\bf 1458}, 482-510 (2000).  
\bibitem{berg} H.C. Berg, {\it E. coli in Motion}, (Springer, 2003).
\bibitem{heuvel} M.G.L. van den Heuvel and C. Dekker, {\it Motor proteins 
at work for nanotechnology}, Science {\bf 317}(5836), 333-336 (2007).
\end{thebibliography}
\end{document}